\documentclass[runningheads]{llncs}
\usepackage[T1]{fontenc}
\usepackage{graphicx}
\usepackage{booktabs}
\usepackage{amsmath}
\usepackage{amssymb}
\usepackage{mathrsfs}
\usepackage{multirow}
\usepackage[colorlinks]{hyperref}
\usepackage{color}
\usepackage[misc]{ifsym}

\begin{document}
\title{mmFormer: Multimodal Medical Transformer for Incomplete Multimodal Learning of Brain Tumor Segmentation}
%
%
\author{
Yao Zhang\inst{1,2}$^\star$ \and
Nanjun He\inst{3}$^\star$ \and
Jiawei Yang\inst{4} \and
Yuexiang Li\inst{3} \and
Dong Wei\inst{3} \and
Yawen Huang\inst{3}\Letter \and
Yang Zhang\inst{5} \and
Zhiqiang He\inst{5}\Letter \and
Yefeng Zheng\inst{3}
}
%
\authorrunning{Yao Zhang et al.}
%
\institute{Institute of Computing Technology, Chinese Academy of Sciences, Beijing, China \and University of Chinese Academy of Sciences, Beijing, China \and Jarvis Lab, Tencent, Shenzhen, China \and Electrical and Computer Engineering, University of California, Los Angeles, USA \and Lenovo Research, Beijing, China \\
\email{zhangyao215@mails.ucas.ac.cn}}
\renewcommand{\thefootnote}{}
%
\maketitle              

\footnote{$^\star$: equal contribution. This work is done when Yao Zhang is an intern at Jarvis Lab, Tencent. Yawen Huang and Zhiqiang He are the corresponding authors.}

\begin{abstract}
Accurate brain tumor segmentation from Magnetic Resonance Imaging (MRI) is desirable to joint learning of multimodal images. However, in clinical practice, it is not always possible to acquire a complete set of MRIs, and the problem of missing modalities causes severe performance degradation in existing multimodal segmentation methods. In this work, we present \textit{the first attempt} to exploit the Transformer for multimodal brain tumor segmentation that is robust to any combinatorial subset of available modalities. Concretely, we propose a novel multi\textbf{m}odal \textbf{M}edical Trans\textbf{former} (\textbf{mmFormer}) for \textit{incomplete multimodal learning} with three main components: the hybrid modality-specific encoders that bridge a convolutional encoder and an intra-modal Transformer for both local and global context modeling within each modality; an inter-modal Transformer to build and align the long-range correlations across modalities for modality-invariant features with global semantics corresponding to tumor region; a decoder that performs a progressive up-sampling and fusion with the modality-invariant features to generate robust segmentation. Besides, auxiliary regularizers are introduced in both encoder and decoder to further enhance the model's robustness to incomplete modalities. We conduct extensive experiments on the public BraTS $2018$ dataset for brain tumor segmentation. The results demonstrate that the proposed mmFormer outperforms the state-of-the-art methods for incomplete multimodal brain tumor segmentation on almost all subsets of incomplete modalities, especially by an average 19.07\% improvement of Dice on tumor segmentation with only one available modality. The code is available at \href{https://github.com/YaoZhang93/mmFormer}{https://github.com/YaoZhang93/mmFormer}.

\end{abstract}
\section{Introduction}
\label{sec:intro}

Automated and accurate segmentation of brain tumors plays an essential role in clinical assessment and diagnosis. Magnetic Resonance Imaging (MRI) is a common neuroimaging technique for the quantitative evaluation of brain tumors in clinical practice, where multiple imaging modalities, i.e., T1-weighted (T1), contrast-enhanced T1-weighted (T1c), T2-weighted (T2), and Fluid Attenuated Inversion Recovery (FLAIR) images, are provided. Each imaging modality provides a distinctive contrast of the brain structure and pathology. The joint learning of multimodal images for brain tumor segmentation is essential and can significantly boost the segmentation performance. Plenty of methods have been widely explored to effectively fuse multimodal MRIs for brain tumor segmentation by, for example, concatenating multimodal images in channel dimension as the input or fusing features in the latent space~\cite{zhou2018one,tseng2017joint}. 
However, in clinical practice, it is not always possible to acquire a complete set of MRIs due to data corruption, various scanning protocols, and unsuitable conditions of patients. In this situation, most existing multimodal methods may fail to deal with incomplete imaging modalities and face a severe degradation in segmentation performance. Consequently, a robust multimodal method is highly desired for a flexible and practical clinical application with one or more missing modalities.

\textit{Incomplete multimodal learning}, also known as hetero-modal learning~\cite{havaei2016hemis}, aims at designing methods that are robust with any subset of available modalities at inference. A straightforward strategy for incomplete multimodal learning of brain tumor segmentation is synthesizing the missing modalities by generative models~\cite{tulder2015does}. Another stream of methods explores knowledge distillation from complete modalities to incomplete ones~\cite{chen2021learning,hu2020knowledge,wang2021acn}. Although promising results are obtained, such methods have to train and deploy a specific model for each subset of missing modalities, which is complicated and burdensome in clinical application. Zhang et al.~\cite{zhang2021modality} proposed an ensemble learning of single-modal models with adaptive fusion to achieve multimodal segmentation. However, it only works when one or all modalities are available. Meanwhile, all these methods require complete modalities during the training process. 

Recent methods focused on learning a unified model, instead of a bunch of distilled networks, for incomplete multimodal segmentation~\cite{havaei2016hemis,shen2019brain}. For example, HeMIS~\cite{havaei2016hemis} learns an embedding of multimodal information by computing mean and variance across features from any number of available modalities. U-HVED~\cite{dorent2019hetero} further introduces multimodal variational auto-encoder to benefit incomplete multimodal segmentation with generation of missing modalities. More recent methods also proposed to exploit feature disentanglement~\cite{chen2019robust} and attention mechanism~\cite{ding2021rfnet} for robust multimodal brain tumor segmentation. Fully Convolutional Network (FCN)~\cite{long2015fully,ronneberger2015u} has achieved great success in medical image segmentation and is widely used for feature extraction in the methods mentioned above. Despite its excellent performance, the inductive bias of convolution, i.e., the locality, makes FCN difficult to build long-range dependencies explicitly. In incomplete multimodal learning of brain tumor segmentation, the features extracted with limited receptive fields tend to be biased when dealing with varying modalities. In contrast, a modality-invariant embedding with global semantic information of tumor region across different modalities may contribute to more robust segmentation, especially when one or more modalities are missing. 

Transformer was originally proposed to model long-range dependencies for sequence-to-sequence tasks~\cite{vaswani2017attention}, and also shows state-of-the-art performance on various computer vision tasks~\cite{dosovitskiy2020image}. Concurrent works~\cite{wang2021transbts,hatamizadeh2022unetr,peiris2021volumetric} exploited Transformer for brain tumor segmentation from the view of backbone network. However, the dedicated Transformer for multimodal modeling of brain tumor segmentation has not been carefully tapped yet, letting alone the incomplete multimodal segmentation. 

\begin{figure}[!tp]
\centering
\includegraphics[width=\linewidth]{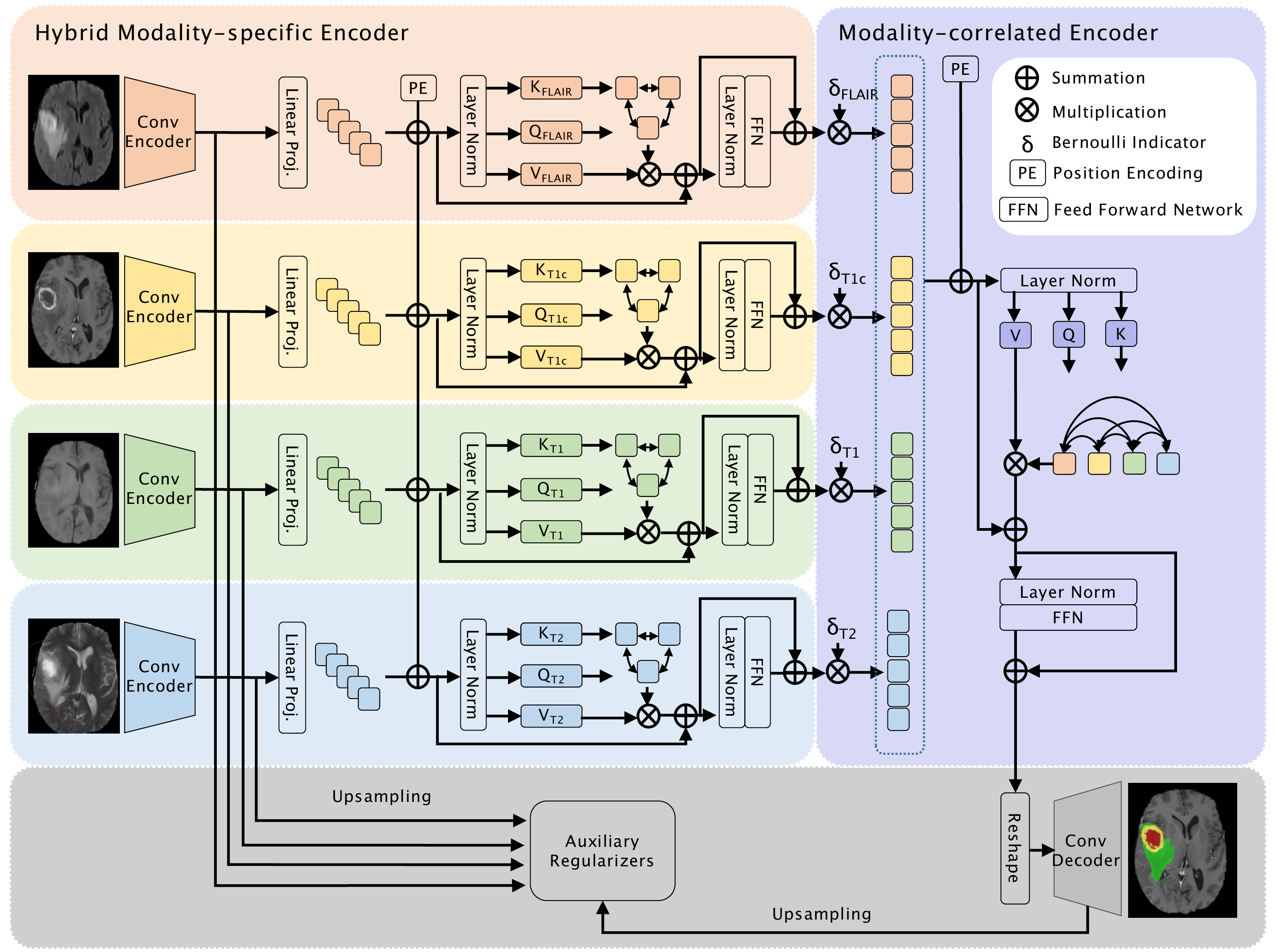}	
\caption{Overview of the proposed mmFormer, which is composed of four hybrid modality-specific encoders, a modality-correlated encoder, and a convolutional decoder. Meanwhile, auxiliary regularizers are introduced in both encoder and decoder. The skip connections between the convolutional encoder and decoder are hidden for clear display.
}
\label{fig:overview}
\end{figure}

This paper aims to exploit Transformer to build a unified model for incomplete multimodal learning of brain tumor segmentation. We propose Multi\textbf{m}odal \textbf{M}edical Trans\textbf{former} (\textbf{mmFormer}) that leverages hybrid modality-specific encoders and a modality-correlated encoder to build the long-range dependencies both within and across different modalities. With the modality-invariant representations extracted by explicitly building and aligning global correlations between different modalities, the proposed mmFormer demonstrates superior robustness to incomplete multimodal learning of brain tumor segmentation. Meanwhile, auxiliary regularizers are introduced into mmFormer to encourage both encoder and decoder to learn discriminative features even when a certain number of modalities are missing. We validate mmFormer on the task of multimodal brain tumor segmentation with BraTS $2018$ dataset~\cite{menze2014multimodal}. The proposed method outperforms the state-of-the-art methods in the average Dice metric over all settings of missing modalities, especially by an average 19.07\% improvement in Dice on enhancing tumor segmentation with only one available modality. To the best of our knowledge, \textit{this is the first attempt to involve the Transformer for incomplete multimodal learning of brain tumor segmentation}.

\section{Method}
\label{sec:method}

In this paper, we propose mmFormer for incomplete multimodal learning of brain tumor segmentation. We adopt an encoder-decoder architecture to construct our mmFormer, including a hybrid modality-specific encoder for each modality, a modality-correlated encoder, and a convolutional decoder. Besides, auxiliary regularizers are introduced in both encoder and decoder. An overview of mmFormer is illustrated in Fig.~\ref{fig:overview}. We elaborate on the details of each component in the followings.


\subsection{Hybrid Modality-specific Encoder.} 
The hybrid modality-specific encoder aims to extract both local and global context information within a specific modality by bridging a convolutional encoder and an intra-modal Transformer. We denote the complete set of modalities by $M=\{FLAIR, T1c, T1, T2\}$. Given an input of $\mathbf{X}_m \in \mathbb{R}^{1\times D\times H\times W}$ with a size of $D\times H\times W$, $m\in M$, we first utilize the convolutional encoder to generate compact feature maps with the local context and then leverage the intra-modal Transformer to model the long-range dependency in a global space.

\noindent\textbf{Convolutional Encoder.}
The convolutional encoder is constructed by stacking convolutional blocks, similar to the encoder part of U-Net~\cite{ronneberger2015u}. The feature maps with the local context within each modality produced by the convolutional encoder $\mathcal{F}^{conv}_m$ can be formulated as
\begin{equation}
	\begin{aligned}
	    \mathbf{F}^{local}_m = \mathcal{F}^{conv}_m(\mathbf{X}_m; \theta^{conv}_m)
	\end{aligned}
\end{equation}
where $\mathbf{F}^{local}_m \in \mathbb{R}^{C\times \frac{D}{2^{l-1}}\times \frac{H}{2^{l-1}}\times \frac{W}{2^{l-1}}}$, $C$ is the channel dimension, and $l$ is the number of the stages in the encoder. Concretely, we build a five-stage encoder, and each stage consists of two convolutional blocks. Each block contains cascaded group normalization, ReLU, and convolutional layers with kernel size of $3$, while the first convolutional block in the first stage only contains a convolutional layer. Between two consecutive blocks, a convolutional layer with stride of $2$ is employed to downsample the feature maps. The number of filters at each level of the encoder is 16, 32, 64, 128, and 256, respectively.

\noindent\textbf{Intra-modal Transformer.}
Limited by the intrinsic locality of the convolutional network, the convolutional encoder fails to effectively build the long-range dependency within each modality. Therefore, we exploit the Intra-modal Transformer for explicitly long-range contextual modeling. The Intra-modal Transformer contains a tokenizer, a Multi-head Self Attention (MSA), and a Feed-Forward Network (FFN). 

As Transformer processes the embeddings in a sequence-to-sequence manner, the local feature maps $\mathbf{F}^{local}_m$ produced by the convolutional encoder is first flattened into a $1$D sequence and transformed into token space by a linear projection. However, the flattening operation inevitably collapses the spatial information, which is critical to image segmentation. To address this issue, we introduce a learnable position embedding $\mathbf{P_m}$ to supplement the flattened features via element-wise summation, which is formulated as
\begin{equation}
	\begin{aligned}
	    \mathbf{F}^{token}_m = \mathbf{F}^{local}_m\mathbf{W}_m + \mathbf{P}_m,
	\end{aligned}
\end{equation}
where $\mathbf{F}^{token}_m \in \mathbb{R}^{C' \times \frac{DHW}{2^{3(l-1)}}}$ denotes the token and $\mathbf{W}_m$ denotes the weights of linear projection. The MSA builds the relationship within each modality by looking over all possible locations in the feature map, which is formulated as
\begin{equation}
	\begin{aligned}
	    head^i_m = Attention(\mathbf{Q}^i_m,\mathbf{K}^i_m,\mathbf{V}^i_m) = softmax(\frac{\mathbf{Q}^i_m\mathbf{K}^{i\mathrm{T}}_m}{\sqrt{d_k}})\mathbf{V}^i_m,
	\end{aligned}
\end{equation}
\begin{equation}
	\begin{aligned}
	    MSA_m = [head^1_m, ..., head^N_m]\mathbf{W}^o_m,
	\end{aligned}
\end{equation}
where $\mathbf{Q}^i_m=LN(\mathbf{F}^{token}_m)\mathbf{W}^{Qi}_m$, $\mathbf{K}^i_m=LN(\mathbf{F}^{token}_m)\mathbf{W}^{Ki}_m$,  $\mathbf{V}^i_m=LN(\mathbf{F}^{token}_m)\mathbf{W}^{Vi}_m$, $LN(\cdot)$ is layer normalization, $d_k$ is the dimension of $\mathbf{K}_m$, $N=8$ is the number of attention heads, and $[\cdot, \cdot]$ is a concatenation operation. The FFN is a two-layer perceptron with GELU~\cite{hendrycks2016gaussian} activation. The feature maps with global context within each modality produced by the intra-modal Transformer is defined as
\begin{equation}
	\begin{aligned}
	    \mathbf{F}^{global}_m = FFN_m(LN(z)) + z, z = MSA_m(LN(\mathbf{F}^{token}_m)) + \mathbf{F}^{token}_m,
	\end{aligned}
\end{equation}
where $\mathbf{F}^{global}_m \in \mathbb{R}^{C' \times \frac{DHW}{2^{3(l-1)}}}$.

\subsection{Modality-correlated Encoder}
The modality-correlated encoder is designed to build the long-range correlations across modalities for modality-invariant features with global semantics corresponding to the tumor region. It is implemented as an inter-modal Transformer. 
\noindent\textbf{Inter-modal Transformer.} In contrast to the intra-modal Transformer, the inter-modal Transformer combines the embeddings from all modality-specific encoders by concatenation as the input multimodal token, which is defined as
\begin{equation}
	\begin{aligned}
	    \mathbf{F}^{token} = [\delta_{FLAIR}\mathbf{F}^{global}_{FLAIR}, \delta_{T1c}\mathbf{F}^{global}_{T1c}, \delta_{T1}\mathbf{F}^{global}_{T1}, \delta_{T2}\mathbf{F}^{global}_{T2}]\mathbf{W} + \mathbf{P},
	\end{aligned}
\end{equation}
where $\delta_m \in \{0, 1\}$ is a Bernoulli indicator that aims to grant robustness when building long-range dependencies between different modalities even when some modalities are missing. This kind of modality-level dropout is randomly conducted during training by setting $\delta_m$ to 0. In case of missing modalities, the multimodal token for the missing modalities will be held by a zero vector.
Subsequently, it is processed by $MSD$ and $FFN$ for modality-invariant features across modalities, which is formulated as
\begin{equation}
	\begin{aligned}
	    \mathbf{F}^{global} = FFN(LN(z)) + z, z = MSA(LN(\mathbf{F}^{token})) + \mathbf{F}^{token},
	\end{aligned}
\end{equation}
where $\mathbf{F}^{global} \in \mathbb{R}^{C' \times \frac{DHW}{2^{(l-1)}}}$.

\subsection{Convolutional Decoder}
The convolutional decoder is designed to progressively restore the spatial resolution from high-level latent space to orinial mask space. The output sequence $\mathbf{F}^{global}$ of the modality-correlated Transformer is reshaped into feature maps corresponding to the size before flattening. The convolutional decoder has a symmetric architecture of convolutional encoder, similar to U-Net~\cite{ronneberger2015u}. Besides, the skip connections between encoder and decoder are also added to keep more low-level details for better segmentation. The features from convolutional encoders of different modalities at a specific level are concatenated and forwarded as skip features to the convolutional decoder.

\subsection{Auxiliary Regularizer}
Conventional multimodal learning models tend to recognize brain tumors relying on the discriminative modalities~\cite{chen2019robust,ding2021rfnet}. Such models are likely to face severe degradation when the discriminative modalities are missing. Therefore, it is critical to encourage each convolutional encoder to segment brain tumors even without the assistance of other modalities. To this end, the outputs of convolutional encoders are upsampled by a shared-weight decoder to segment tumors from each modality separately. The shared-weight decoder has the same architecture with the convolutional decoder. 
Besides, we also introduce auxiliary regularizers in the convolutional decoder to force the decoder to generate accurate segmentation even when certain modalities are missing. It is achieved by interpolating the feature maps in each stage of the convolutional decoder to segment tumors via deep supervision~\cite{dou20173d}. 
Dice loss~\cite{milletari2016v} is employed as the regularizer. Combining the training loss of the network's output with the auxiliary regularizers, the overall loss function is defined as
\begin{equation}
\label{eq:dice}
\begin{aligned}
	\mathcal{L}=1-Dice=1-\frac{2 \sum_{c=1}^{C} \sum_{i=1}^{N_c}  g_{i}^{c} p_{i}^{c}}{\sum_{c=1}^{C} \sum_{i=1}^{N_c} g_{i}^{c 2}+\sum_{c=1}^{C} \sum_{i=1}^{N_c} p_{i}^{c 2}},
\end{aligned}
\end{equation}
\begin{equation}
	\begin{aligned}
	    \mathcal{L}_{\text {total }} = \sum_{i\in M}\mathcal{L}^{encoder}_i + \sum^{l-1}_{i=1}\mathcal{L}^{decoder}_i + \mathcal{L}^{output},
	\end{aligned}
\end{equation}
where $C$ is the number of segmentation classes, and $N_c$ is the number of voxels of class $c$, $g_i^c$ is a binary indicator if class label $c$ is the correct classification for pixel $i$, $p_i^c$ is the corresponding predicted probability, $M=\{FLAIR, T1c, T1, T2\}$, and $l$ is the number of stages in the convolutional decoder.

\section{Experiments and Results}
\label{sec:exp}
\subsubsection{Dataset and Implementation.} The experiments are conducted on BraTS $2018$ dataset\footnote{https://www.med.upenn.edu/sbia/brats2018/data.html}~\cite{menze2014multimodal}, which consists of 285 multi-contrast MRI scans with four modalities: T1, T1c, T2, and FLAIR. 
Different subregions of brain tumors are combined into three nested subregions: whole tumor, tumor core, and enhancing tumor. All the volumes have been co-registered to the same anatomical template and interpolated to the same resolution by the organizers. Dice Similarity Coefficient (DSC) as defined in Eq. (\ref{eq:dice}) is employed for evaluation.
The framework is implemented with PyTorch $1.7$ on four NVIDIA Tesla V$100$ GPUs. The input size is $128 \times 128 \times 128$ voxels and batch size is $1$. Random flip, crop, and intensity shifts are employed for data augmentation. The mmFormer has $106$M parameters and $748$G FLOPs. The network is trained with the Adam optimizer with an initial learning rate of $0.0002$ for $1000$ epochs. The model is trained for about $25$ hours with $17$G memory on each GPU. 

\begin{table*}[!tp]
\centering
\caption{Results of the proposed method and state-of-the-art unified models, i.e., HeMIS~\cite{havaei2016hemis} and U-HVED~\cite{dorent2019hetero}, on BraTS 2018 dataset~\cite{menze2014multimodal}. Dice similarity coefficient (DSC) [\%] is employed for evaluation with every combination settings of modalities. $\bullet$ and $\circ$ denote available and missing modalities, respectively.}
\label{tab:incomplete}
\resizebox{\textwidth}{!}{
\begin{tabular}{cccc|ccc|ccc|ccc}
\toprule
\multicolumn{4}{c|}{Modalities}  & \multicolumn{3}{c|}{Enhancing Tumor} & \multicolumn{3}{c|}{Tumor Core} & \multicolumn{3}{c}{Whole Tumor}\\ \hline
F &T1c &T1 &T2 & U-HeMIS & U-HVED & Ours & U-HeMIS & U-HVED & Ours & U-HeMIS & U-HVED & Ours\\ \midrule \midrule
$\bullet$ &$\circ$ &$\circ$ &$\circ$               & 11.78     & 23.80     &  \textbf{39.33} & 26.06  & 57.90 & \textbf{61.21} & 52.48  & 84.39 & \textbf{86.10}   \\ \hline
$\circ$ &$\bullet$ &$\circ$ &$\circ$               & 62.02        & 57.64     & \textbf{72.60} & 65.29  & 59.59 & \textbf{75.41}  & 61.53  & 53.62 & \textbf{72.22}     \\ \hline
$\circ$ &$\circ$ &$\bullet$ &$\circ$               & 10.16        & 8.60     & \textbf{32.53}  & 37.39    & 33.90  & \textbf{56.55}  & 57.62 & 49.51 & \textbf{67.52}   \\ \hline
$\circ$ &$\circ$ &$\circ$ &$\bullet$               & 25.63        & 22.82     & \textbf{43.05}  & 57.20  & 54.67 & \textbf{64.20}  & 80.96 & 79.83 & \textbf{81.15}    \\ \hline
$\bullet$ &$\bullet$ &$\circ$ &$\circ$               & 66.10        & 68.36     & \textbf{75.07}  & 71.49 & 75.07  & \textbf{77.88} & 68.99  & 85.93  & \textbf{87.30}     \\ \hline
$\bullet$ &$\circ$ &$\bullet$ &$\circ$               & 10.71        & 27.96     & \textbf{42.96} & 41.12  & 61.14 & \textbf{65.91} & 64.62  & 85.71  & \textbf{87.06}      \\ \hline
$\bullet$ &$\circ$ &$\circ$ &$\bullet$               & 30.22        & 32.31     & \textbf{47.52}  & 57.68 & 62.70  & \textbf{69.75} & 82.95 & 87.58   & \textbf{87.59}      \\ \hline
$\circ$ &$\bullet$ &$\bullet$ &$\circ$               & 66.22        & 61.11     & \textbf{74.04}  & 72.46    & 67.55     &  \textbf{78.59}   & 68.47 & 64.22 & \textbf{74.42}      \\ \hline
$\circ$ &$\bullet$ &$\circ$ &$\bullet$               & 67.83        & 67.83     & \textbf{74.51} & 76.64  & 73.92 & \textbf{78.61}  & 82.48   & 81.32  & \textbf{82.99}     \\ \hline
$\circ$ &$\circ$ &$\bullet$ &$\bullet$               & 32.39        & 24.29     & \textbf{44.99} & 60.92  & 56.26 & \textbf{69.42}   & \textbf{82.41}  & 81.56  & 82.20       \\ \hline
$\bullet$ &$\bullet$ &$\bullet$ &$\circ$               & 68.54        & 68.60     & \textbf{75.47} & 76.01  & 77.05 & \textbf{79.80}  & 72.31 & 86.72  & \textbf{87.33}       \\ \hline
$\bullet$ &$\bullet$ &$\circ$ &$\bullet$               & 68.72        & 68.93     & \textbf{75.67} & 77.53  & 76.75 & \textbf{79.55}  & 83.85  & 88.09  & \textbf{88.14}       \\ \hline
$\bullet$ &$\circ$ &$\bullet$ &$\bullet$               & 31.07        & 32.34     & \textbf{47.70} & 60.32  & 63.14 & \textbf{71.52}  & 83.43  & \textbf{88.07}  & 87.75       \\ \hline
$\circ$ &$\bullet$ &$\bullet$ &$\bullet$               & 69.92        & 67.75     & \textbf{74.75}  & 78.96 & 75.28 & \textbf{80.39}  & \textbf{83.94}  & 82.32  & 82.71      \\ \hline
$\bullet$ &$\bullet$ &$\bullet$ &$\bullet$        & 70.24       & 69.03   & \textbf{77.61}     & 79.48  & 77.71  & \textbf{85.78}  & 84.74  & 88.46  & \textbf{89.64}      \\ \hline
\multicolumn{4}{c|}{Average}        & 46.10       & 46.76   & \textbf{59.85}     & 62.57  & 64.84  & \textbf{72.97}  & 74.05  & 79.16  & \textbf{82.94}      \\ \bottomrule
\end{tabular}%
}
\end{table*}

\begin{figure}[!tp]
\centering
\includegraphics[width=\linewidth]{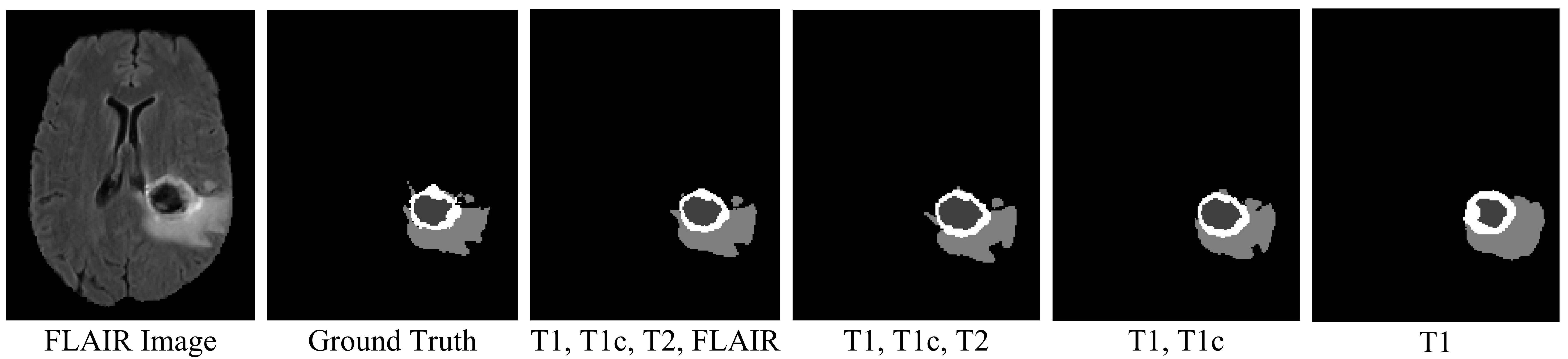}	
\vspace{-6mm}
\caption{Segmentation results of mmFormer with various available modalities.}
\vspace{-4mm}
\label{fig:incomplete}
\end{figure}

\noindent\textbf{Performance of Incomplete Multimodal Segmentation.}
We evaluate the robustness of our method to incomplete multimodal segmentation. The absence of modality is implemented by setting $\delta_i, i\in \{FLAIR, T1c, T1, T2\}$ to be zero for dropping the specific modalities at inference.
We compare our method with two representative models using shared latent space, i.e., HeMIS~\cite{havaei2016hemis} and U-HVED~\cite{dorent2019hetero}.
For a fair comparison, we use the same data split in~\cite{wang2021acn} and directly reference the results. 
In Table~\ref{tab:incomplete}, our method significantly outperforms HeMIS and U-HVED on the segmentation of enhancing tumor and tumor core on all the $15$ possible combinantions of available modalities and the segmentation of the whole tumor on $12$ out of $15$. 
In Table~\ref{tab:improvement}, we show that with the increased number of missing modalities, the average improvement obtained by mmFormer is more considerable. Meanwhile, it is observed that mmFormer gains more improvement when the target is more difficult to segment. These results demonstrate the effectiveness of mmFormer for incomplete multimodal learning of brain tumor segmentation.
Fig.~\ref{fig:incomplete} shows that even with one modality available, mmFormer can achieve proper segmentation for brain tumor.

We also compare mmFormer with ACN~\cite{wang2021acn}. ACN relies on knowledge distillation for incomplete multimodal brain tumor segmentation. In the case of $N$ modalities in total, ACN has to train $2^4-2$ times to distill $2^N-2$ student models for all conditions of missing modalities, while our mmFormer only learns once by a unified model. Specifically, ACN is trained for $672$ hours with $144$M parameters for $1$ teacher and $14$ student models, while mmFormer requires only $25$ hours with $106$M parameters. Nevertheless, the average DSC for enhancing tumor, tumor core, and whole tumor of mmFormer (59.85, 72.97 and 82.94, respectively) is still close to it of ACN (61.21, 77.62, and 85.92, respectively). 

\noindent\textbf{Performance of Complete Multimodal Segmentation.}
We compare our method with a recent Transformer-based method, i.e., TransBTS~\cite{wang2021transbts}, for multimodal brain tumor segmentation with full modalities. We reproduce the results with the official repository. TransBTS obtains DSC of 72.66\%, 72.69\%, and 79.99\% on enhancing tumor, tumor core, and the whole tumor, respectively. Our mmFormer outperforms TransBTS on all subregions of brain tumor with DSC of 77.61\%, 85.78\%, and 89.64\%, demonstrating the effectiveness of mmFormer even for complete multimodal brain tumor segmentation.

\noindent\textbf{Ablation Study.}
We investigate the effectiveness of intra-modal Transformer, inter-modal Transformer, and auxiliary regularizer as three critical components in our method. We analyze the effectiveness of each component by excluding one of them from mmFormer. In Table~\ref{tab:ablation}, we compare the performance of the three variants to mmFormer with DSC, averaging over the 15 possible combinations of input modalities. It shows that intra-modal Transformer, inter-modal Transformer, and auxiliary regularizer bring performance improvement across all the tumor subregions.


\begin{center}
\begin{minipage}[!tp]{\textwidth}
\begin{minipage}[!tp]{0.48\textwidth}
\centering
\makeatletter\def\@captype{table}
\caption{Average improvements of mmFormer upon HeMIS~\cite{havaei2016hemis} and U-HVED~\cite{dorent2019hetero} with different numbers of missing modalities evaluated by DSC [\%].}
\resizebox{\textwidth}{!}{
\begin{tabular}{c|cccc}
\toprule
\multirow{2}{*}{Regions} & \multicolumn{4}{c}{\# of missing modalities}                                \\ \cline{2-5} 
                               & 0 & 1 & 2 & 3 \\ \hline
Enhancing                      & +7.98  & +8.91  & +13.57 & +19.07  \\ \hline
Core                           & +7.19  & +4.68  & +8.62  & +15.34  \\ \hline
Whole                          & +3.04  & +2.89  & +5.57  & +11.75  \\ \bottomrule
\end{tabular}
}
\label{tab:improvement}
\end{minipage}
\begin{minipage}[!tp]{0.48\textwidth}
\centering
\makeatletter\def\@captype{table}
\caption{Ablation study of critical components of mmFormer.}
\begin{tabular}{l|ccc}
\toprule
\multirow{2}{*}{Methods} & \multicolumn{3}{c}{Average DSC [\%]}                               \\ \cline{2-4} 
                         & Enhancing & Core & Whole \\ \midrule
mmFormer                 &  59.85  & 72.97  & 82.94   \\ \hline
w/o IntraTrans                 &  56.98  & 71.83   &  81.32  \\ \hline
w/o InterTrans                 &  56.05  & 70.28   &  81.12  \\ \hline
w/o Aux. Reg.                 &  55.78  & 69.33   &  81.65  \\ \bottomrule
\end{tabular}
\label{tab:ablation}
\end{minipage}
\end{minipage}
\end{center}

\section{Conclusion}
\label{sec:concolusion}

We proposed a Transformer-based method for incomplete multimodal learning of brain tumor segmentation. The proposed mmFormer bridges Transformer and CNN to build the long-range dependencies both within and across different modalities of MRI images for a modality-invariant representation. We validated our method on brain tumor segmentation under various combinations of missing modalities, and it outperformed state-of-the-art methods on the BraTS benchmark. Our method gains more improvements when more modalities are missing and/or the target ones are more difficult to segment.

%
%
%

\bibliographystyle{splncs04}
\bibliography{paper363}
%
\end{document}